\pdfoutput=1

\documentclass[11pt,twoside,a4paper,cmspaper,final,collab]{cms-tdr}
\def\svnVersion{395171}\def\cmsCernNoTag{CERN-EP/2016-265}\def\cmsCernDate{\today}\def\cmsMessage{Published in the European Physical Journal C as \href{http://dx.doi.org/10.1140/epjc/s10052-017-4718-8}{\doi{10.1140/epjc/s10052-017-4718-8.}}}

\begin{document}\cmsNoteHeader{TOP-16-005}

\hyphenation{had-ron-i-za-tion}
\hyphenation{cal-or-i-me-ter}
\hyphenation{de-vices}

\RCS$Revision: 386187 $
\RCS$HeadURL: svn+ssh://svn.cern.ch/reps/tdr2/papers/TOP-16-005/trunk/TOP-16-005.tex $
\RCS$Id: TOP-16-005.tex 386187 2017-02-09 15:59:23Z palencia $
\newlength\cmsFigWidth
\ifthenelse{\boolean{cms@external}}{\setlength\cmsFigWidth{0.85\columnwidth}}{\setlength\cmsFigWidth{0.4\textwidth}}
\ifthenelse{\boolean{cms@external}}{\providecommand{\cmsLeft}{top\xspace}}{\providecommand{\cmsLeft}{left\xspace}}
\ifthenelse{\boolean{cms@external}}{\providecommand{\cmsRight}{bottom\xspace}}{\providecommand{\cmsRight}{right\xspace}}

\providecommand{\eepm}{\ensuremath{\Pep\Pem}\xspace}
\providecommand{\mmpm}{\ensuremath{\Pgmp \Pgmm}\xspace}
\providecommand{\empm}{\ensuremath{\Pe^\pm \Pgm^\mp}\xspace}
\providecommand{\mll}{\ensuremath{M_{\ell\ell}}\xspace}

\newcommand{\amcatnlo}   {\textsc{mg5}\_a\MCATNLO\xspace}
\newcommand{\madspin}   {\textsc{madspin}\xspace}
\newcommand{\xseccalctoterr}{\ensuremath{\sigma_{\ttbar} = 832^{+40}_{-46}\unit{pb}}\xspace}
\newcommand{\trigeff}{\ensuremath{0.99 \pm 0.01}\xspace}
\newcommand{\ideffmu}{90\%}
\newcommand{\ideffele}{80\%}
\newcommand{\sfmu}{0.98}
\newcommand{\sfele}{0.99}

\cmsNoteHeader{TOP-16-005}
\title{Measurement of the $\ttbar$ production cross section using events in the $\Pe \PGm$ final state in pp collisions at $\sqrt{s}=13\TeV$ }

\titlerunning{The ttbbar production cross section using events in the e mu final state in pp collisions at 13 TeV}

\date{\today}

\newcommand{\xsec}{\ensuremath{\sigma_{\ttbar} = 815 \pm 9\stat \pm 38\syst \pm 19\lum\unit{pb}}\xspace}
\newcommand{\xsecee}{\ensuremath{\sigma_{\ttbar} = 767 \pm 17\stat \pm 56\syst \pm 21\lum\unit{pb}}\xspace}
\newcommand{\xsecmm}{\ensuremath{\sigma_{\ttbar} = 824 \pm 14\stat \pm 47\syst \pm 22\lum\unit{pb}}\xspace}
\newcommand{\xsecfit}{\ensuremath{\sigma_{\ttbar} = 815 \pm 7\stat \pm 33\syst\pm 49\lum\unit{pb}}\xspace}
\newcommand{\fidxsec}{\ensuremath{\sigma_{\ttbar} = 14.3 \pm 0.1 \stat \pm 0.7 \syst\pm 0.4 \lum\unit{pb}}\xspace}
\newcommand{\usedLumi}{2.2\fbinv}
\newcommand{\reluncertainty}{5.3}
\newcommand{\totuncertainty}{43.0}

\abstract{
The cross section of top quark-antiquark pair production in proton-proton collisions at $\sqrt{s} = 13\TeV$ is measured by the CMS experiment at the LHC, using data corresponding to an integrated luminosity of 2.2\fbinv. The measurement is performed by analyzing events in which the final state includes one electron, one muon, and two or more jets, at least one of which is identified as originating from hadronization of a b quark. The  measured cross section is $815 \pm 9\stat\pm 38\syst \pm 19\lum\unit{pb}$, in agreement with the expectation from the standard model.
}

\hypersetup{%
pdfauthor={CMS Collaboration},%
pdftitle={Measurement of the ttbbar production cross section using events in the e mu final state in pp collisions at sqrt(s)=13 TeV},%
pdfsubject={CMS},%
pdfkeywords={CMS, physics, top physics}}

\maketitle

\section{Introduction}
\label{sec:introduction}
The measurement of the top quark-antiquark pair (\ttbar) cross section provides a test of the hadro\-production
of top quark pairs as predicted by quantum chromodynamics (QCD). At the CERN LHC, measurements have been performed
in many different decay channels and at three different proton-proton collision energies~\cite{CMStt13,CMStt12,CMStt11,
CMStt10,CMStt9,CMStt8,CMStt6,CMStt5,CMStt4,CMStt3,CMStt2,top15003,topPAS13004,ATLAStt14,ATLAStt13,ATLAStt12,ATLAStt11,
ATLAStt10,ATLAStt9,ATLAStt8,ATLAStt6,ATLAStt3,ATLAStt1, ATLAStt13TeV2}. Precision measurements of these cross sections
allow for a test of their energy dependence as predicted by QCD; they can also place constrains on the parton distribution
functions (PDFs)~\cite{Czakon:2013tha}. In combination with some theory, they also provide unambiguous
measurements of interesting quantities, such as the top quark pole mass~\cite{ATLAStt6, topPAS13004}, which is difficult to
determine by other means. A detailed understanding of the production cross section is also required in searches for evidence
of new physics beyond the standard model, as \ttbar production is often the dominant background process. This is especially
important if the signature for the new physics is similar to that of \ttbar production~\cite{Aad:2015pfx, topPAS13004}. This
paper presents a measurement of the \ttbar production cross section ($\sigma_{\ttbar}$) in the \empm decay channel using
an event-counting method, based on observed yields. The analysis follows closely~\cite{top15003}, and uses the full data set recorded
by CMS at 13\TeV during 2015, which corresponds to an integrated luminosity of \usedLumi. This represents a factor of 50 increase in the amount of data over the original analysis and
allows for more detailed studies of the experimental and theory uncertainties.

\section{The CMS detector and Monte Carlo simulation}
\label{sec:detector}

The CMS detector~\cite{Chatrchyan:2008zzk} has a superconducting solenoid
in its central region that provides an axial magnetic field
of 3.8\unit{T}. The silicon pixel and strip trackers cover
$0 < \phi <2\pi$ in azimuth and $\abs{\eta}<2.5$ in pseudorapidity.
The lead tungstate crystal electromagnetic calorimeter, and the brass and scintillator
hadron calorimeter are located inside the solenoid. These are used to identify electrons,
photons and jets. Muons are measured in gas-ionization detectors embedded in the steel
flux-return yoke outside the solenoid. The detector is nearly hermetic, providing reliable
measurement of the momentum imbalance in the plane transverse to the beams. A two-level
trigger system selects the most interesting $\Pp\Pp$ collisions for offline analysis.
A more detailed description of the CMS detector, together with a definition
of the coordinate system used and the relevant kinematic variables, can be
found in Ref.~\cite{Chatrchyan:2008zzk}.

{\tolerance=800
Different Monte Carlo (MC) event generators are used to simulate signal and background
events. The next-to-leading-order (NLO) \POWHEG~(v2)~\cite{powheg,powheg2} generator is used for
\ttbar events, with the top quark mass ($m_{\cPqt}$) set to 172.5\GeV. The NNPDF3.0
NLO~\cite{nnpdf} PDFs are used.
For the reference \ttbar sample, the events are interfaced
with \PYTHIA~(v8.205) \cite{Sjostrand:2006za,Sjostrand:2014zea}
with the CUETP8M1 tune~\cite{CMS-PAS-GEN-14-001, Skands:2014pea}
to simulate parton showering, hadronization, and  the underlying
event. Additional samples are produced by showering the events in the reference sample
with \HERWIGpp ~(v2.7.1)~\cite{herwigpp} or by generating events using
\amcatnlo~(v5\_2.2.2)~\cite{Alwall:2014hca} interfaced with \madspin ~\cite{madspin} to account for
spin correlations in the decays of the top quarks, and using
\PYTHIA for parton showering and hadronization.
\par}

The \amcatnlo\ generator is also used
to simulate \PW+jets events and Drell--Yan (DY) quark-antiquark
annihilation into lepton-antilepton pairs through a virtual photon
or a Z boson exchange; for these backgrounds the event yields are estimated from data.
Single top quark events are simulated using
\POWHEG~(v1)~\cite{powheg1,powheg3} and \PYTHIA, and the event yields are normalized to
the approximate next-to-next-to-leading order (NNLO) cross
sections from Ref.~\cite{Kidonakis:2013zqa}. The diagram removal approach~\cite{Frixione:2008yi} is used to handle the interference between the \ttbar and tW final states starting at NLO. The contributions
from \PW\PW, \PW\cPZ, and \cPZ\cPZ\ (referred to as ``VV'') processes
are simulated with \PYTHIA, and the event rates are normalized to the NLO cross sections from Ref.~\cite{mcfm}.
Other contributions from \PW\ and \cPZ\ boson production in association with \ttbar events
(referred to as ``$\ttbar$V'') are simulated using \amcatnlo~and \PYTHIA.
The simulated samples include additional interactions per bunch crossing (pileup), with the distribution matching that observed in data,
with an average of about 11 collisions per bunch crossing.

{\tolerance=1200
The SM prediction for $\sigma_{\ttbar}$
at 13\TeV is $832^{+20}_{-29}\,\text{(scales)}\pm 35\,$(PDF$+\alpha_s)\unit{pb}$
for $m_{\cPqt}=172.5\GeV$, as calculated with the \textsc{Top++} program~\cite{top++} at
NNLO in perturbative QCD, including
soft-gluon resummation at next-to-next-to-leading-log
order~\cite{mitov}.
The first uncertainty reflects uncertainties in the
factorization ($\mu_\mathrm{F}$) and renormalization ($\mu_\mathrm{R}$) scales.
The second one is associated with possible choices of PDFs and the value of the strong coupling constant,
following the PDF4LHC prescriptions~\cite{pdf4lhcReport, pdf4lhcInterim}, using the MSTW2008 68\% confidence level
NNLO~\cite{Martin:2009iq,mstw08}, CT10 NNLO~\cite{Lai:2010vv,pdfsets},
and NNPDF2.3 5f FFN~\cite{Ball:2012cx} PDF sets.
The expected event yields for signal in all figures and tables are normalized to this cross section.
\par}

\section{Event selection}
\label{sec:event_selection_yields}

In the SM, top quarks in $\Pp\Pp$  collisions are mostly produced as \ttbar pairs, where
each top quark decays predominantly to a \PW\ boson and a bottom quark.
In \ttbar events where both \PW\ bosons decay leptonically, the final state contains two leptons
of opposite electric charge and at least two jets coming from the hadronization of the bottom quarks.

At the trigger level, a combination of the single lepton and dilepton triggers is used. Events are
required to contain either one electron with transverse momentum $\pt > 12\GeV$
and one muon with $\pt > 17\GeV$ or one electron with $\pt> 17\GeV$ and one muon with $\pt > 8\GeV$.
In addition, single-lepton triggers with one electron (muon) with $\pt > 23\GeV$\,(20) are used in order to increase the efficiency.
The efficiency for the combination of the single lepton and dilepton triggers is measured in data using
triggers based on \pt imbalance in the event. The trigger efficiency is measured
to be \trigeff\ (combined statistical and systematic uncertainties) when the selection on the leptons described below is applied. The
trigger in simulation is corrected using a multiplicative data-to-simulation scale
factor (SF), given by the trigger efficiency measured in data with independent monitoring triggers.

The particle-flow (PF) event algorithm~\cite{CMS:2009nxa, CMS:2010byl} reconstructs and identifies each individual particle
with an optimized combination of information from the various elements of the CMS detector. Selected dilepton events are required to contain one isolated electron~\cite{Khachatryan:2015hwa} and one isolated
muon~\cite{Chatrchyan:2012xi} with opposite electric charge and $\pt > 20\GeV$ and $\abs{\eta} < 2.4$.
Isolation requirements are based on
the ratio of the scalar sum of the transverse momenta of all PF candidates, reconstructed inside a
cone centered on the lepton, excluding the contribution from the lepton candidate.
This isolation variable is required to be smaller than 7\%\,(15\%) of the electron (muon) \pt.

In events with more than one pair of leptons passing the selection, the two
opposite-sign different-flavour leptons with the largest \pt are selected for further study.
Events with \PW\ bosons decaying into \Pgt\ leptons contribute to the measurement
only if the \Pgt\ leptons decay into electrons or muons that
satisfy the selection requirements.

The efficiency of the lepton selection is measured using a ``tag-and-probe"~\cite{Khachatryan:2010xn} method
in a sample of same-flavour dilepton events, which is enriched in Z boson candidates.
The measured \pt- and $\eta$-dependent values for the combined identification and isolation
efficiencies average to about \ideffele\ for electrons and \ideffmu\ for muons.
To account for the difference in efficiencies determined using data and simulation, the event yield in simulation is corrected using
\pt- and $\eta$-dependent SFs based on a comparison of lepton selection efficiencies
in data and simulation. These have an average of \sfele\ for electrons and \sfmu\ for muons.

In order to suppress backgrounds from DY production of $\tau$ lepton pairs
with low invariant dilepton mass, \ttbar candidate events are further required to have
a dilepton pair of invariant mass $m_{\Pe\PGm} > 20\GeV$.

Jets are reconstructed from the PF particle candidates using the anti-$k_t$ clustering algorithm~\cite{Cacciari:2008gp, Cacciari:2011ma} with a distance
parameter of 0.4. The jet momentum is determined from the vectorial sum of all particle momenta in the jet, and is found from simulation to be within 5 to 10\% of the true momentum over the whole \pt spectrum and detector acceptance. An offset correction is applied to jet energies to take into account the contribution from additional proton-proton interactions within the same or nearby bunch crossings. Jet energy corrections are derived from simulation, confirmed with in situ measurements of the energy balance in dijet and photon + jet events, and
are applied as a function of the jet
\pt and $\eta$~\cite{Khachatryan:2016kdb} to both data and simulated events. The \ttbar candidate events
are required to have at least
two reconstructed jets with $\pt> 30\GeV$ and $\abs{\eta}< 2.4$.

Since \ttbar events decay into final states containing a bottom quark-antiquark pair,
requiring the presence of jets identified as originating
from \cPqb\ quarks (``\cPqb\ jets'') reduces backgrounds from DY and \PW+jets production.
Jets are identified as \cPqb\ jets using the combined secondary vertex algorithm
~\cite{Chatrchyan:2012jua, BTV15001}, with an operating point which yields an identification efficiency of 67\% and a  misidentification (mistag)
probability of about 1\% and 15\%~\cite{BTV15001} for light-flavour jets (\cPqu, \cPqd, \cPqs,
and gluons) and \cPqc\ jets, respectively. The selection requires the presence of at least one \cPqb\ jet in the event.

\section{Background determination}
\label{sec:bkgd}

Background events arise primarily from single top quark, DY, and VV events
in which at least two prompt leptons are produced by Z or \PW\ boson decays. The single
top quark and VV contributions are estimated from simulation.

The DY event yield is estimated from data using the ``$R_\text{out/in}$''
method~\cite{CMStt8,CMStt12,CMStt13}, where events with same-flavour leptons
are used to normalize the yield of \empm\ pairs from DY production of $\tau$ lepton pairs. A data-to-simulation normalization
factor is estimated from the number of events in data
within a 15\GeV window around the Z boson mass
and extrapolated to the number of events outside the Z mass window
with corrections applied using control regions enriched in DY events
in data. The SF is found to be $0.95 \pm 0.05$ (statistical uncertainty) after applying the final event selection.

Other background sources, such as \ttbar or \PW+jets events in the lepton+jets final state,
can contaminate the signal sample if a jet is incorrectly
reconstructed as a lepton, or the lepton is incorrectly identified as being isolated. This is
more important for electrons. For muons, the dominant contribution comes from the semileptonic
decay of bottom or charm quarks. These events are grouped into the nonprompt leptons category (``non-\PW/Z
leptons'') since prompt leptons are defined as originating from decays of \PW\ or Z
boson, together with contributions that can arise, for example, from
decays of mesons or photon conversions.

The contribution of non-\PW/Z lepton events is estimated from a control region
of same-sign (SS) events and propagated in the opposite-sign (OS) signal region.
The SS control region is defined using the same criteria as the
nominal signal region, except for requiring \Pe\Pgm\ pairs with the same electric charge.
The SS dilepton events are predominantly events containing misidentified leptons.
Other SM processes
produce prompt SS or charge-misidentified dilepton events with significantly smaller rates;
these are estimated using simulation and subtracted from the observed number of
events in data.

The scaling from the SS control region in data to the signal region is
performed through the ratio of the numbers of OS to SS events with misidentified leptons in simulation.
This ratio is calculated using simulated \ttbar and \PW+jets samples, which are rich in
nonprompt dilepton events,
and is measured to be $1.4 \pm 0.1\stat$.
In data, 152 SS events are observed, with a contribution of $79.8\pm 1.9\stat$
prompt lepton SS events as evaluated from simulation. In total $104 \pm 8\,\text{(stat + syst)}$ events with misidentified leptons
contaminating the signal region are predicted. This agrees within the uncertainties with predictions from the
simulation.

\begin{figure}[htbp!]
\centering
{\includegraphics[width=0.49\textwidth]{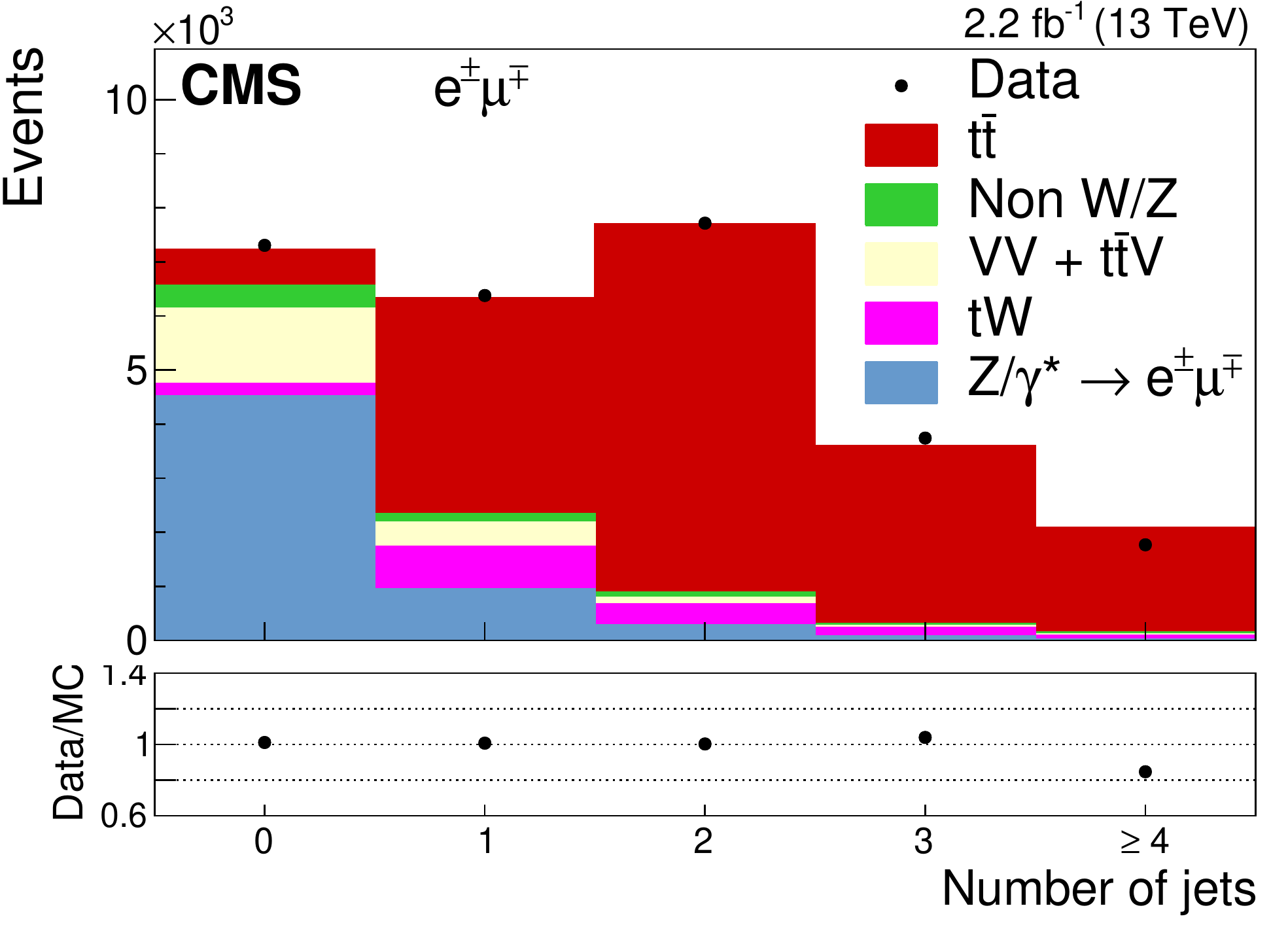}}
\caption{Distribution of the jet multiplicity
in events passing the dilepton selection criteria.
The expected distributions for \ttbar signal and individual backgrounds are shown
after corrections based on control regions in data are applied; the
last bin contains the overflow events.
The ratio of data to the sum of the expected yields is given at the bottom of
the figure. The error bars, which are within the size of the points, indicate the statistical uncertainties.
}
\label{fig:dilepton}
\end{figure}

\begin{figure*}[htbp!]
\centering
{\includegraphics[width=0.49\textwidth]{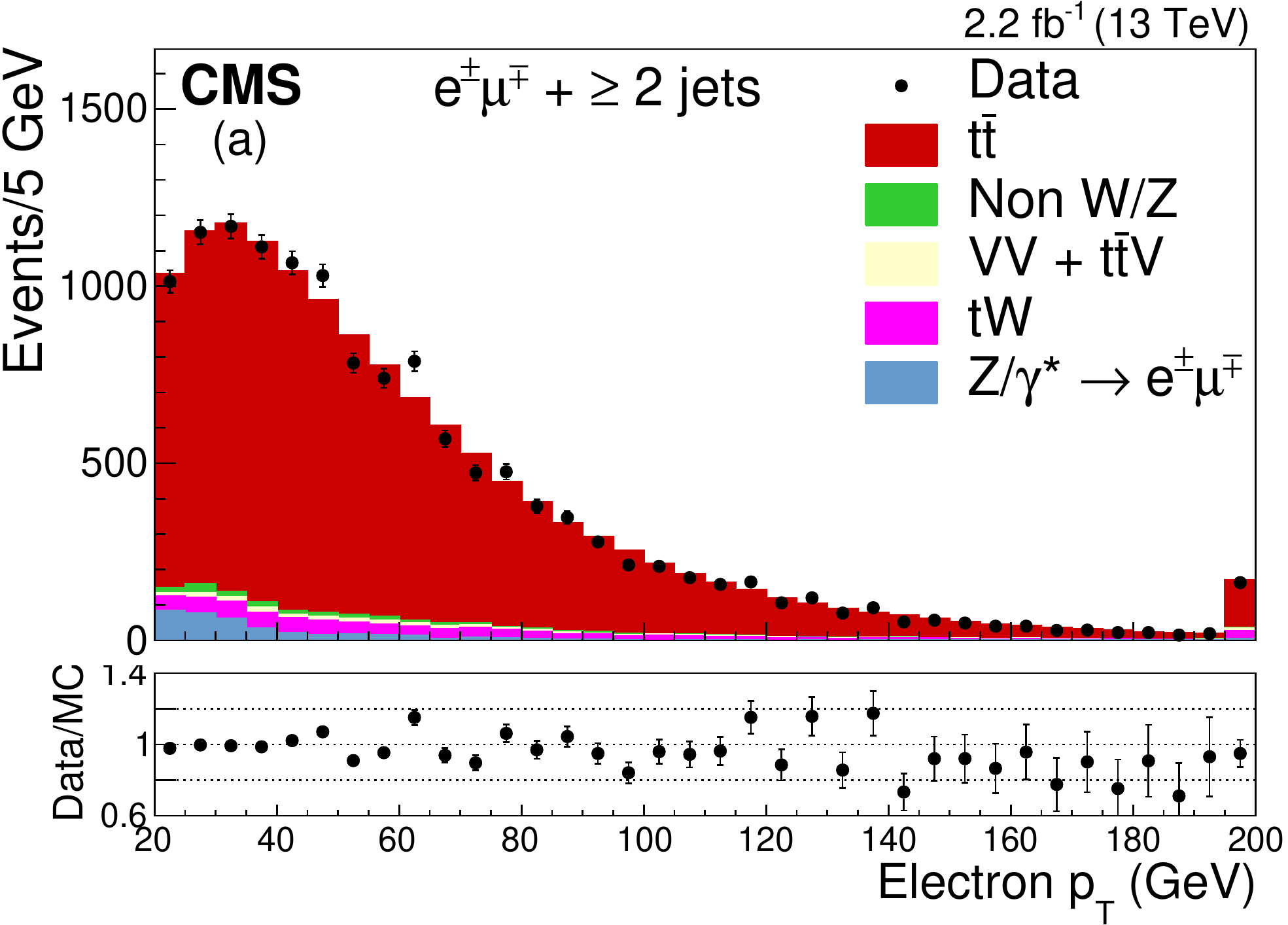}}
{\includegraphics[width=0.49\textwidth]{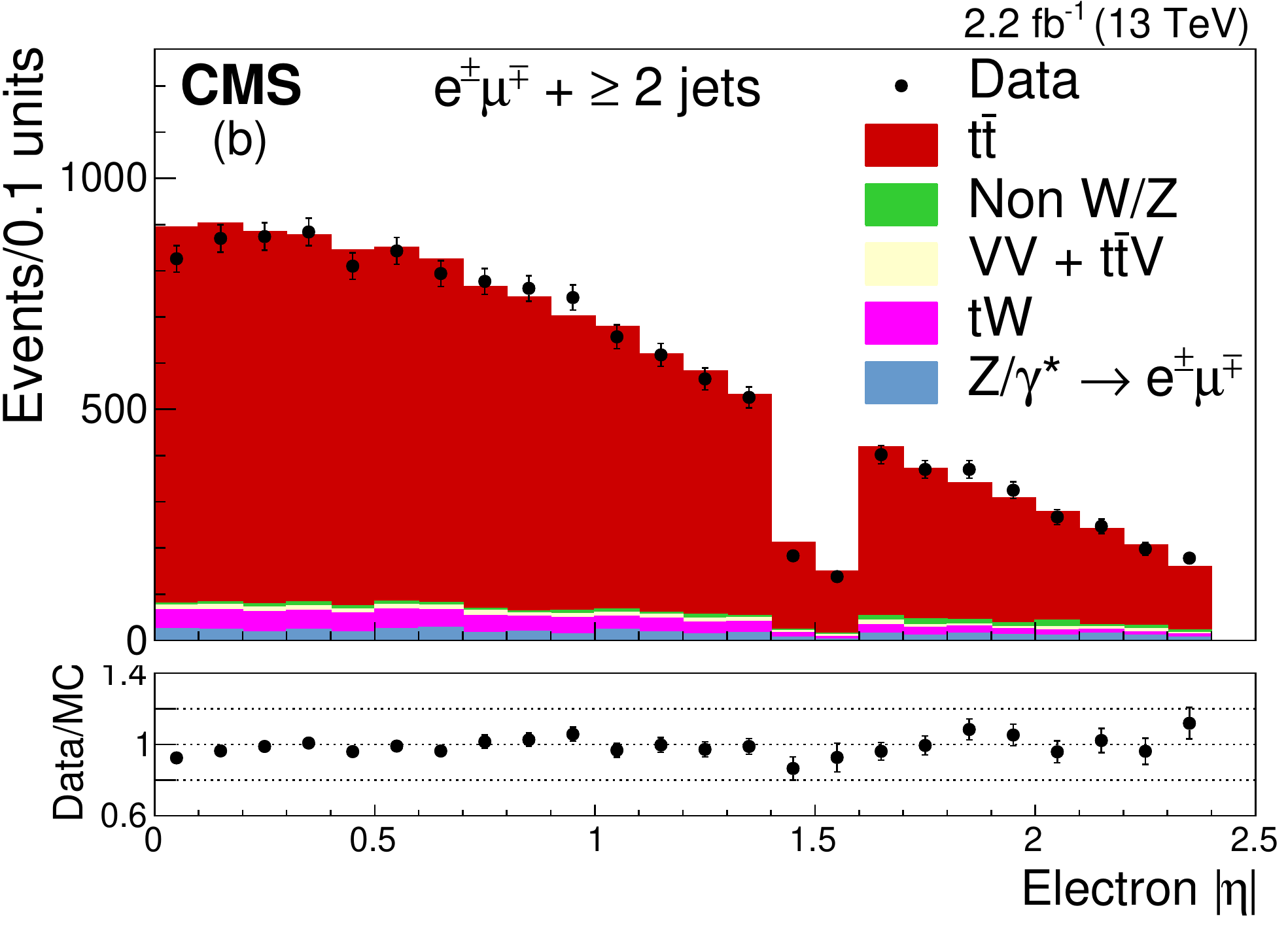}}\\
\vspace{0.1cm}
{\includegraphics[width=0.49\textwidth]{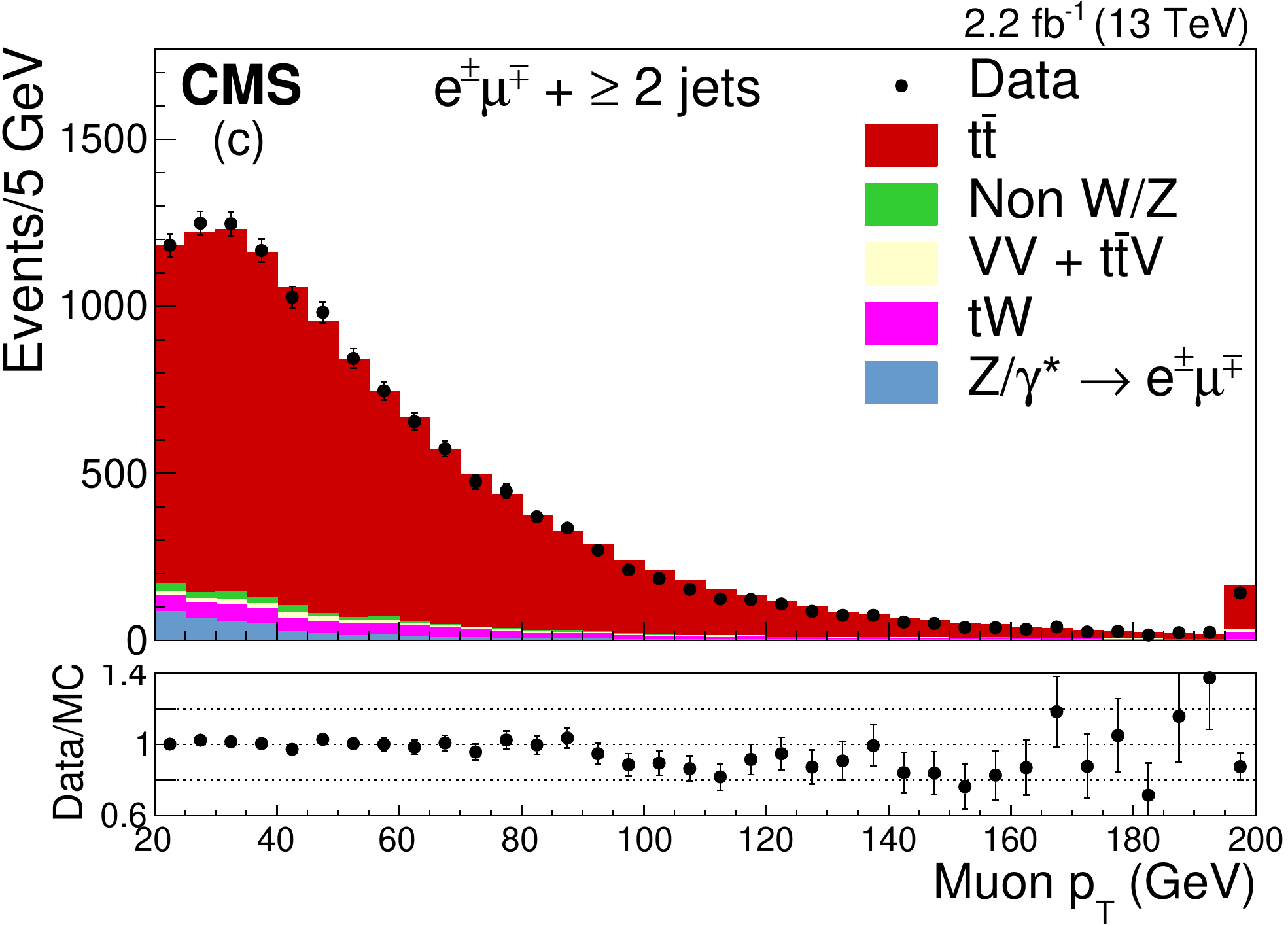}}
{\includegraphics[width=0.49\textwidth]{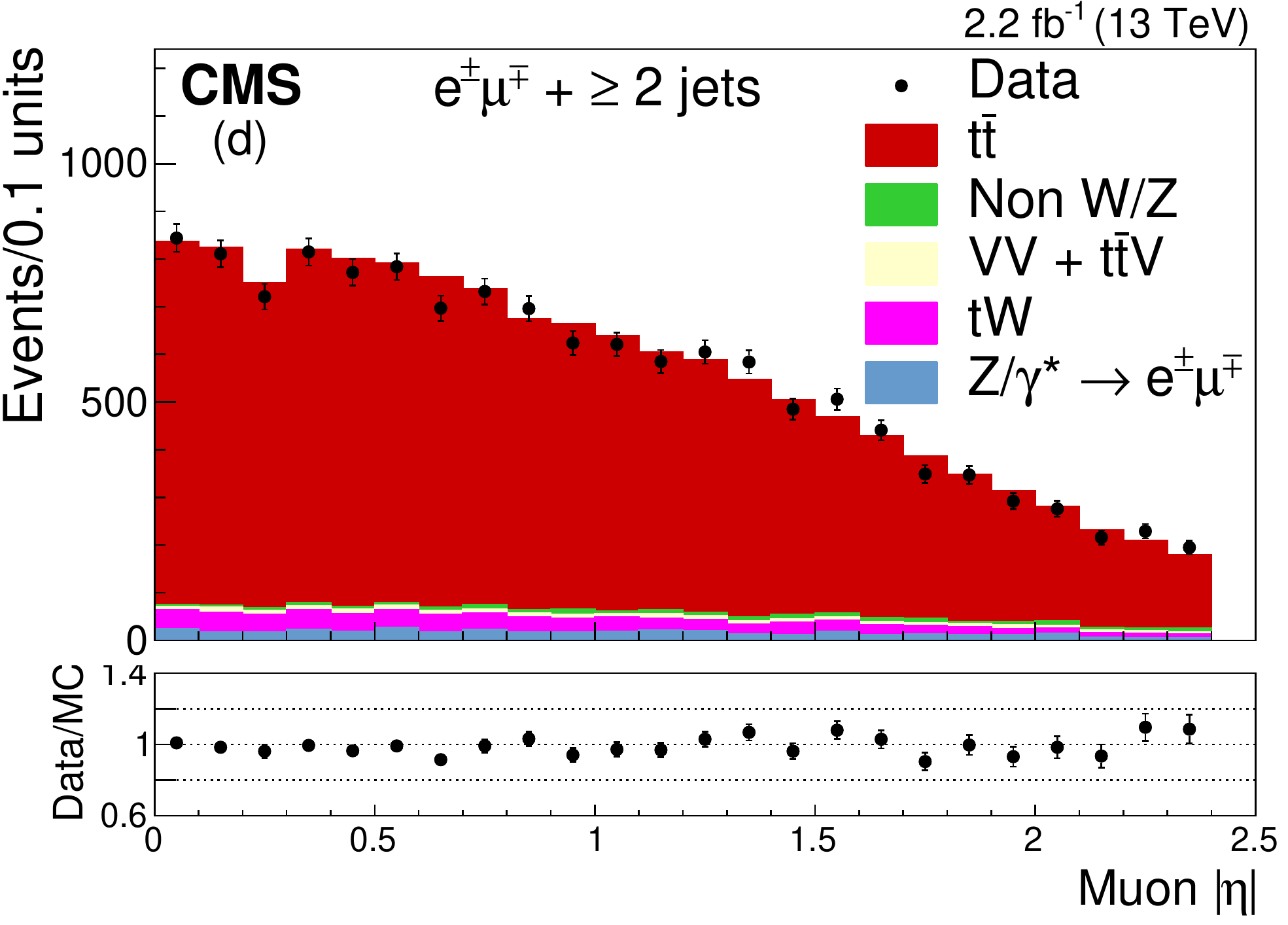}}
\caption{The distributions of (a) \pt and (b) $\abs{\eta}$
of the electron, and (c) \pt and (d) $\abs{\eta}$ of the muon after the selection of jets and before
the \cPqb\ jet requirement. The expected distributions for \ttbar signal and individual
backgrounds are shown after corrections based on control regions in data are applied; for
the left plots (a, c) the last bin contains the overflow events.
The ratios of data to the sum of the expected yields are given at the bottom of each
panel. The error bars indicate the statistical uncertainties.
}
\label{fig:leptons}
\end{figure*}

\begin{figure*}[htbp!]
\centering
{\includegraphics[width=0.49\textwidth]{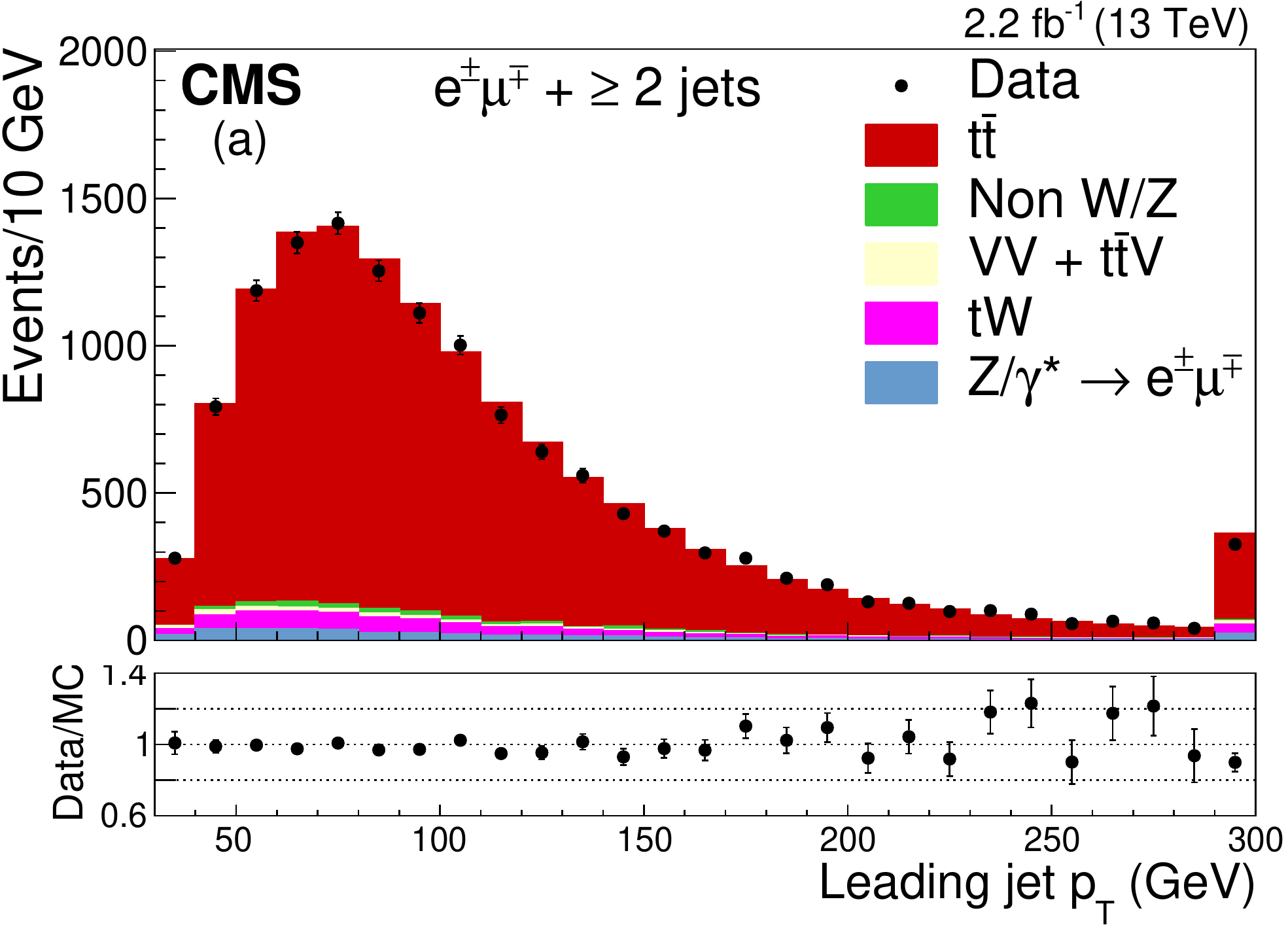}}
{\includegraphics[width=0.49\textwidth]{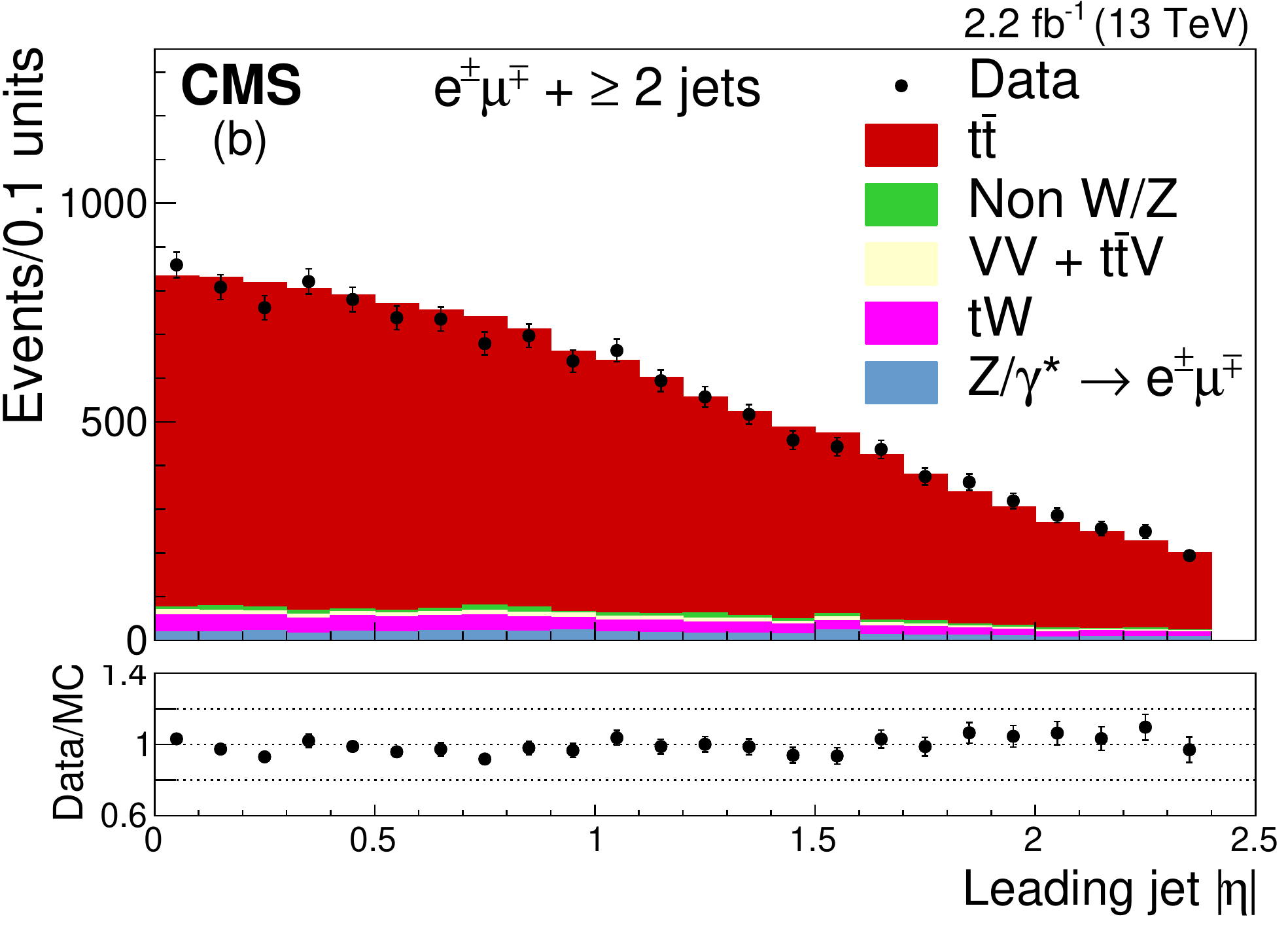}}\\
{\includegraphics[width=0.49\textwidth]{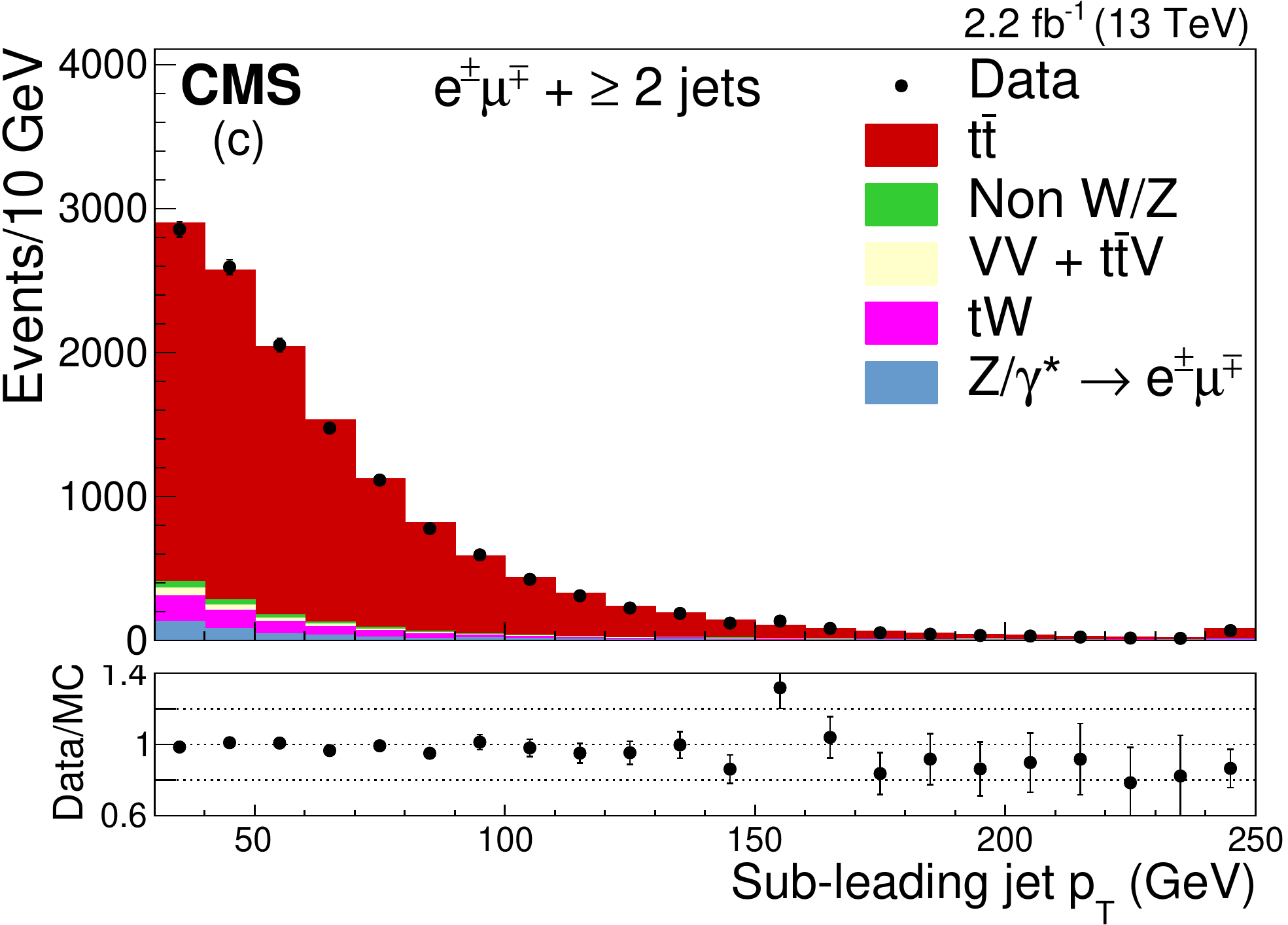}}
{\includegraphics[width=0.49\textwidth]{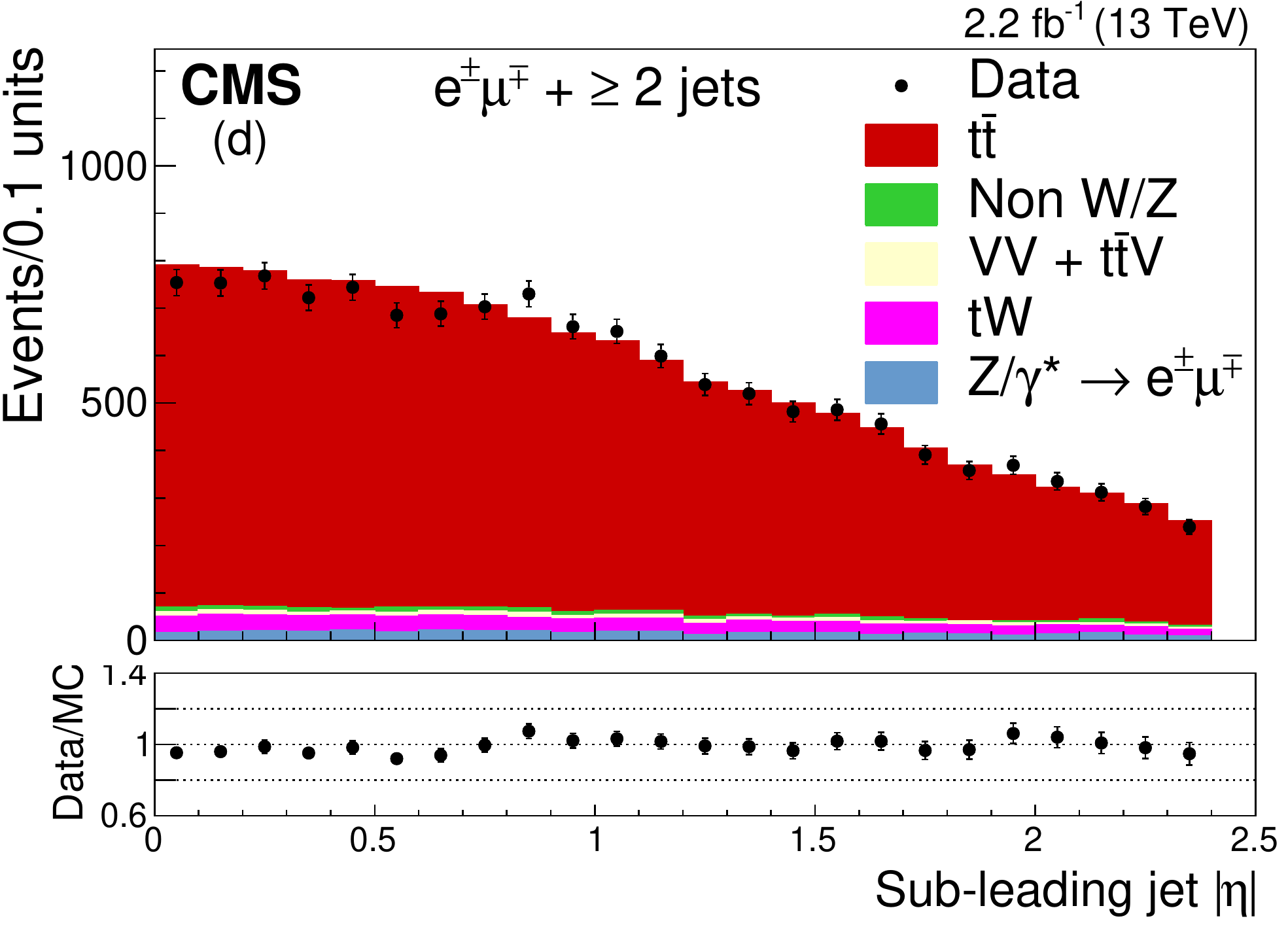}}\\
{\includegraphics[width=0.49\textwidth]{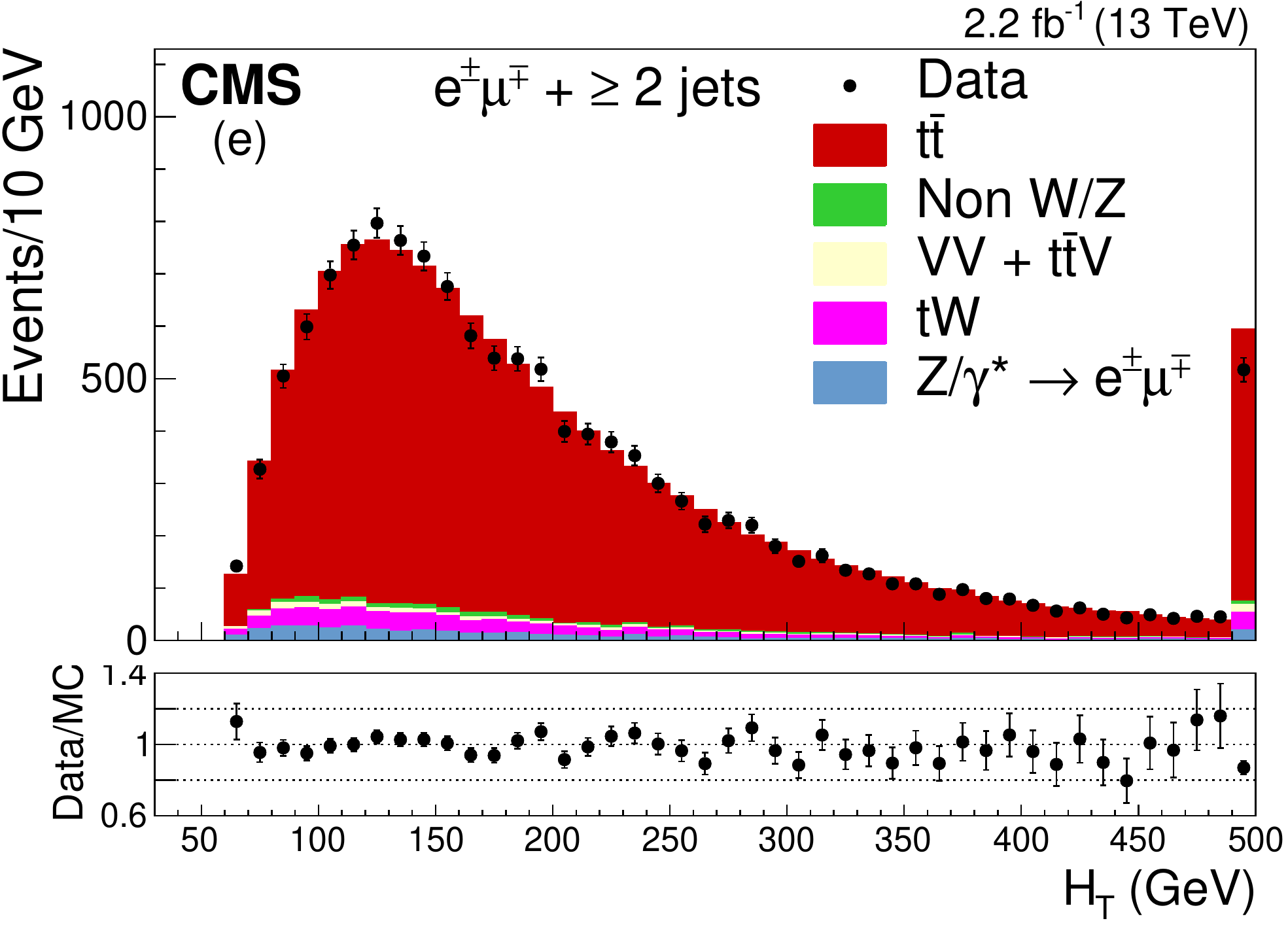}}
{\includegraphics[width=0.49\textwidth]{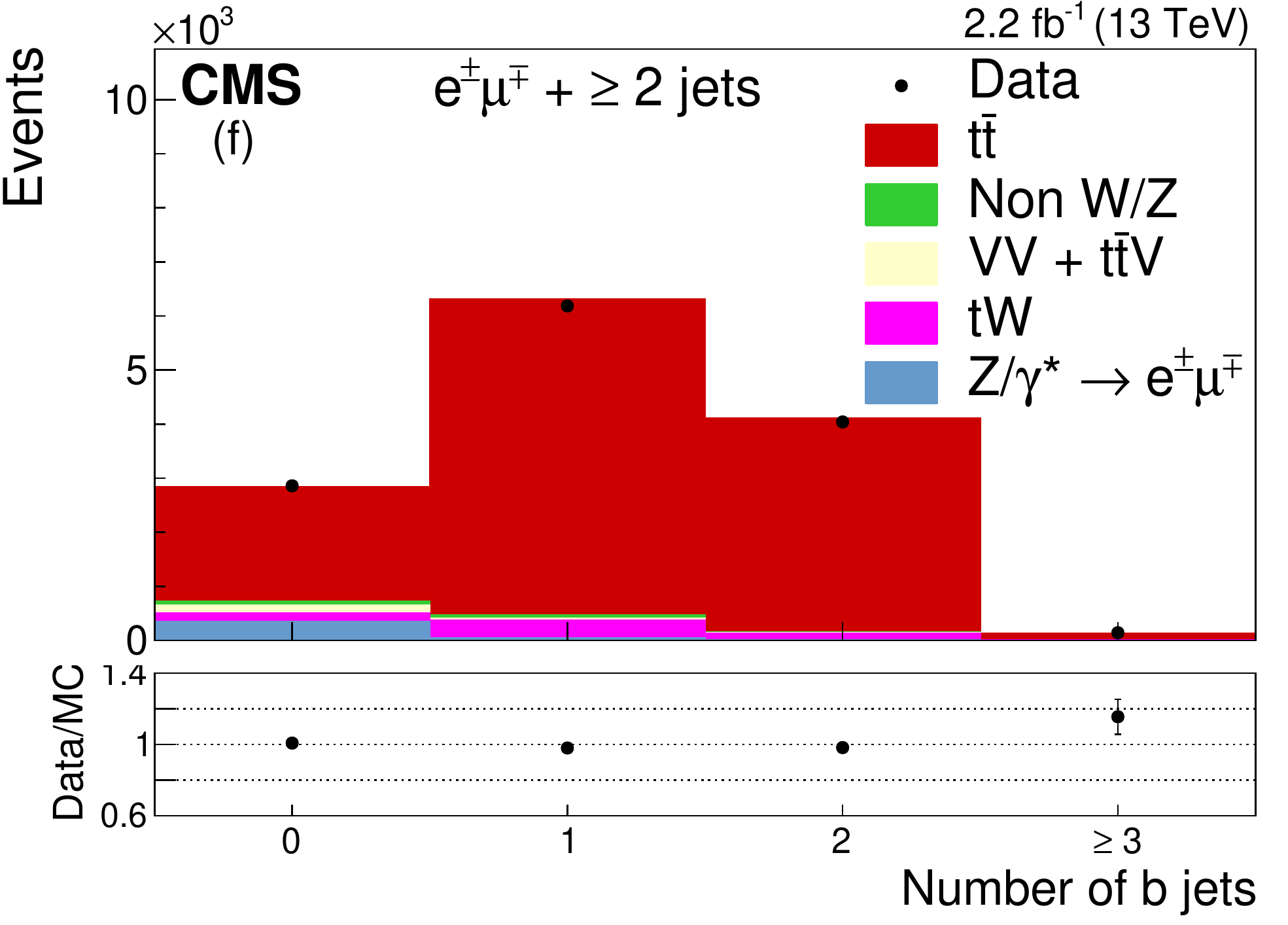}}
\caption{The distributions of (a) \pt and (b) $\abs{\eta}$
for the leading jet, (c) \pt and (d) $\abs{\eta}$ for the
sub-leading jet, (e) $H_\mathrm{T}$, and (f) b jet multiplicity after the jets selection
and before the \cPqb\ jet requirement.
The expected distributions for \ttbar signal and individual
backgrounds are shown after corrections based on control regions in data are applied; in each plot the
last bin contains the overflow events.
The ratios of data to the sum of the expected yields are given at the bottom of each panel.
The error bars indicate the statistical uncertainties.
}
\label{fig:jets}
\end{figure*}

Figure~\ref{fig:dilepton} shows the multiplicity of jets
for events passing the dilepton criteria. The MC simulation does not describe well the data
for events with $\geq$4 jets, the region in which parton shower effects are expected
to dominate the prediction.
After requiring at least two jets, Fig.~\ref{fig:leptons} shows the \pt and
$\abs{\eta}$ distributions of the selected leptons, and
Fig.~\ref{fig:jets} shows the \pt (a, c) and
$\abs{\eta}$ (b, d) distributions of the two most energetic jets; Fig.~\ref{fig:jets} (e) shows the
scalar sum of the transverse momenta of all jets ($H_\mathrm{T}$) and Fig.~\ref{fig:jets} (f) the b jet multiplicity.
Good agreement between data and the predictions for signal and background is observed.

\section{Sources of systematic uncertainty}
\label{sec:systematics}
Table~\ref{tab:breakdown_comb} summarizes the statistical uncertainty and the different
sources of systematic uncertainties in the measured \ttbar production cross section.

\begin{table}[!htb]
\centering
\topcaption{Summary of the individual contributions to the uncertainty in
the $\sigma_{\ttbar}$  measurement.
The first and second uncertainty corresponds to the total and relative component, respectively.
The total uncertainty in the result, calculated as the quadratic sum of the individual components, is also given.}

\begin{tabular}{lcc}
\hline
Source                        &  $\Delta\sigma_{\ttbar}$ (pb)  & $\Delta\sigma_{\ttbar} / \sigma_{\ttbar}$ (\%) \\
\hline
\multicolumn{3}{c}{Experimental}\\
\hline
Trigger efficiencies          &  9.9 & 1.2 \\
Lepton efficiencies           & 18.9 & 2.3 \\
Lepton energy scale           & $<$1 & $\leq$0.1\\
Jet energy scale              & 17.4 & 2.1 \\
Jet energy resolution         &  0.8 & 0.1 \\
b tagging                     & 11.0 & 1.3 \\
Mistagging                    & $<$1 & $\leq$0.1\\
Pileup                        &  1.5 & 0.2 \\
\hline
\multicolumn{3}{c}{Modeling}\\
\hline
$\mu_\mathrm{F}$ and $\mu_\mathrm{R}$ scales & $<$1 & $\leq$0.1 \\
\ttbar NLO generator          & 17.3 & 2.1 \\
\ttbar hadronization          &  6.0 & 0.7 \\
Parton shower scale           &  6.5 & 0.8 \\
PDF                           &  4.9 & 0.6 \\
\hline
\multicolumn{3}{c}{Background}\\
\hline
Single top quark              & 11.8 & 1.5 \\
VV                            & $<$1 & $\leq$0.1\\
Drell--Yan                    & $<$1 & $\leq$0.1 \\
Non-\PW/Z leptons             &  2.6 & 0.3 \\
$\ttbar$V                     & $<$1 & $\leq$0.1 \\
\hline
Total systematic              & \multirow{2}{*}{37.8}  &  \multirow{2}{*}{4.6} \\
(no integrated luminosity)    &      &     \\
Integrated luminosity         & 18.8 & 2.3 \\
Statistical                   &  8.5 & 1.0 \\
\hline
Total                         & \totuncertainty  &  \reluncertainty  \\
\hline
\end{tabular}
\label{tab:breakdown_comb}
\end{table}

The uncertainty in the trigger efficiency SF applied to simulation to correct for
differences with respect to data is 1.1\%. The uncertainty in the SF applied to
correct the electron (muon) identification efficiency is found to be about 1.8\%\,(1.5\%),
with some dependence on the lepton \pt and $\eta$.

The modeling of lepton energy scales was studied using $\Z
\to {\Pe\Pe} / \PGm\PGm$
events in data and simulation, resulting in an uncertainty for the
electron (muon) energy scale of 1.0\,(0.5)\%.
These values are used to obtain the effect on the signal acceptance, which is taken
as a systematic uncertainty.

The impact of uncertainties in jet energy scale (JES) and jet energy
resolution (JER) is estimated from the change observed in the number
of simulated \ttbar events selected after changing the jet momenta within
the JES uncertainties, and for JER by an $\abs{\eta}$-dependent variation
of the JER scale factors within their uncertainties.

The uncertainties resulting from the \cPqb\ tagging efficiency and misidentification rate
are determined by varying the \cPqb\ tagging SF of the \cPqb\ jets and the light-flavour jets, respectively.
These uncertainties depend on the \pt and $\eta$ of the jet and amount to approximately 2\% for \cPqb\ jets and 10\% for
mistagged jets~\cite{BTV15001} in \ttbar signal events. They are propagated to the \ttbar selection
efficiency using simulated events.

The uncertainty assigned to the number of pileup events in simulation is obtained by changing the
inelastic proton-proton cross section, which is used to estimate the pileup in data, by
$\pm$5\%~\cite{Aaboud:2016mmw}.

The systematic uncertainty related to the missing higher-order diagrams
in \POWHEG is estimated as follows: the uncertainty in the signal acceptance
is determined by changing the $\mu_\mathrm{F}$ and $\mu_\mathrm{R}$ scales in
\POWHEG independently up and down by a factor of two,
with the uncertainty taken as the maximum observed difference.

The predictions of the NLO generators \POWHEG and  \amcatnlo\ for
\ttbar production are compared, where both use \PYTHIA for hadronization, fragmentation,  and
additional radiation description. The difference in the signal acceptance between the two is taken as an
uncertainty.

The uncertainty arising from the hadronization model
mainly affects the JES and the  fragmentation
of \cPqb\ quark jets.
The uncertainty in the JES already
contains a contribution from the uncertainty in the hadronization.
In addition, we determine a related uncertainty by comparing
samples
of events generated with \POWHEG, where the hadronization is modeled
with \PYTHIA or \HERWIGpp. In what follows we refer to this difference
as the hadronization uncertainty.

The impact of the choice of the parton shower scale is studied by changing the scale of the parton
shower (initial and final state radiation) by a factor of 2 and 1/2 from its default value. The maximum variation with
respect to the central value of the signal acceptance at particle level~\cite{Khachatryan:2015oqa} for the fiducial volume
of the analysis is taken as an uncertainty.

The uncertainty from the choice of PDF
is determined by reweighting the sample of simulated
\ttbar events according to the NNPDF3.0 PDF sets~\cite{nnpdf}. The
root-mean-square of the distribution is taken as an uncertainty.

Based on recent measurements of the production cross section for single top
quark~\cite{CMStopPublicationstop4, Chatrchyan:2014tua, Chatrchyan:2012zca} and
VV~\cite{CMSWWZZPublication8, CMSWWPublication7,  CMSWWHiggsPublication7,
Khachatryan:2016txa, Khachatryan:2016tgp, ATLASWWPublication,ATLASWZPublication,ATLASZZPublication}
we use an uncertainty of 30\% for these background processes.
For DY production, an uncertainty of 15\%, that covers the difference of the SF at different levels of the selection, is assumed.
A 30\% systematic uncertainty is estimated for the non-\PW/Z lepton background
derived from the uncertainty in the ratio of the numbers of OS to SS events with misidentified leptons in the MC simulation.

The uncertainty in the integrated luminosity is 2.3\%~\cite{lumiPAS15}.

\section{Results}
\label{sec:measure}
The \ttbar production cross section is measured by counting events and applying
the expression
\begin{equation*}
\sigma_{\ttbar} = \frac{N - N_\mathrm{B}}{\mathcal{A} \, {\mathcal{L}}},
\label{eqn:xsecA}
\end{equation*}
where $N$ is the  total number of dilepton events observed in data, $N_\mathrm{B}$ is the
number of estimated background events, $\mathcal{A}$ is the product of the mean acceptance,
the selection efficiency, and the
branching fraction into the \empm final state,
and $\mathcal{L}$ is the integrated luminosity.

Table~\ref{tab:yields} shows
the total number of events observed in data together with the total number of signal
and background events determined from simulation or estimated from
data. The value of $\mathcal{A}$, determined from simulation assuming $m_{\cPqt}= 172.5\GeV$,
 is $(0.55\pm 0.03) \% $, including statistical and systematic uncertainties.
The measured cross section is
\begin{equation*}
\xsec,
\end{equation*}
for a top quark mass of 172.5\GeV.

\begin{table}[!htb]
\centering
\topcaption{Number of dilepton events obtained after applying the full selection.
The results are given for the individual sources of background, \ttbar signal
with a top quark mass of 172.5\GeV  and \xseccalctoterr, and data. The
uncertainties correspond to the statistical component.
}
\begin{tabular}{lc}
\hline
         &  Number of \\
Source   & \empm\ events      \\
\hline
Drell--Yan              &    46 $\pm$   5 $\pm$   7\phantom{0}  \\
Non-\PW/Z leptons       &   104 $\pm$   8 $\pm$  31\phantom{0}  \\
Single top quark        &   452 $\pm$   6 $\pm$ 141             \\
VV                      &    14 $\pm$   2 $\pm$   5\phantom{0}  \\
$\ttbar$V               &    30 $\pm$ 	1 $\pm$	  9\phantom{0}  \\ \hline
Total background        &   646 $\pm$  11 $\pm$ 145             \\ \hline
\ttbar signal          & 9\,921 $\pm$  14 $\pm$ 436\phantom{0}  \\ \hline
Data                    &    10368\phantom{0}                   \\
\hline
\end{tabular}
\label{tab:yields}
\end{table}

As a cross-check, analogous measurements have been performed using independent data samples with
same-flavour leptons in the final state.
The results obtained in the \eepm\ and \mmpm\ channels are consistent with the result in the \empm\ channel. Given their
larger uncertainties, the results are not combined with the main one in the \empm\ channel.

The measured fiducial cross section for \ttbar production with two
leptons (one electron and one muon) in the
range $\pt > 20\GeV$ and $\abs{\eta} < 2.4$, at least two jets with $\pt > 30\GeV$ and
$\abs{\eta} < 2.4$, and at least one b jet is $\sigma^\text{fid}_{\ttbar} = 12.4 \pm 0.1\stat \pm 0.5\syst\pm 0.3\lum\unit{pb}$.

The acceptance has been measured in the range 166.5--178.5\GeV and is parameterized as a linear function of $m_{\cPqt}$.
The cross section varies by 3.7\unit{pb} when the top quark mass changes 0.5\GeV.

\section{Summary}
\label{sec:conclusions}

A measurement of the \ttbar production cross section in proton-proton collisions
at $\sqrt{s} =13\TeV$ is presented for events containing an oppositely charged electron-muon pair,
and two or more jets, of which at least one is tagged as originating from a \cPqb\  quark.
The measurement is performed through an event-counting method based on a data
sample corresponding to an integrated luminosity of \usedLumi.
The measured cross section is
\begin{equation*}
\xsec ,
\end{equation*}
with a total relative uncertainty of \reluncertainty\%.
The measurement, that supersedes~\cite{top15003}, is consistent with recent measurements from the ATLAS~\cite{ATLAStt13TeV2} and CMS~\cite{top15003}
experiments and with the standard model prediction of \xseccalctoterr\ for
a top quark mass of 172.5\GeV.

\begin{acknowledgments}
We congratulate our colleagues in the CERN accelerator departments for the excellent performance of the LHC and thank the technical and administrative staffs at CERN and at other CMS institutes for their contributions to the success of the CMS effort. In addition, we gratefully acknowledge the computing centres and personnel of the Worldwide LHC Computing Grid for delivering so effectively the computing infrastructure essential to our analyses. Finally, we acknowledge the enduring support for the construction and operation of the LHC and the CMS detector provided by the following funding agencies: BMWFW and FWF (Austria); FNRS and FWO (Belgium); CNPq, CAPES, FAPERJ, and FAPESP (Brazil); MES (Bulgaria); CERN; CAS, MoST, and NSFC (China); COLCIENCIAS (Colombia); MSES and CSF (Croatia); RPF (Cyprus); SENESCYT (Ecuador); MoER, ERC IUT, and ERDF (Estonia); Academy of Finland, MEC, and HIP (Finland); CEA and CNRS/IN2P3 (France); BMBF, DFG, and HGF (Germany); GSRT (Greece); OTKA and NIH (Hungary); DAE and DST (India); IPM (Iran); SFI (Ireland); INFN (Italy); MSIP and NRF (Republic of Korea); LAS (Lithuania); MOE and UM (Malaysia); BUAP, CINVESTAV, CONACYT, LNS, SEP, and UASLP-FAI (Mexico); MBIE (New Zealand); PAEC (Pakistan); MSHE and NSC (Poland); FCT (Portugal); JINR (Dubna); MON, RosAtom, RAS, and RFBR (Russia); MESTD (Serbia); SEIDI and CPAN (Spain); Swiss Funding Agencies (Switzerland); MST (Taipei); ThEPCenter, IPST, STAR, and NSTDA (Thailand); TUBITAK and TAEK (Turkey); NASU and SFFR (Ukraine); STFC (United Kingdom); DOE and NSF (USA).

\hyphenation{Rachada-pisek} Individuals have received support from the Marie-Curie programme and the European Research Council and EPLANET (European Union); the Leventis Foundation; the A. P. Sloan Foundation; the Alexander von Humboldt Foundation; the Belgian Federal Science Policy Office; the Fonds pour la Formation \`a la Recherche dans l'Industrie et dans l'Agriculture (FRIA-Belgium); the Agentschap voor Innovatie door Wetenschap en Technologie (IWT-Belgium); the Ministry of Education, Youth and Sports (MEYS) of the Czech Republic; the Council of Science and Industrial Research, India; the HOMING PLUS programme of the Foundation for Polish Science, cofinanced from European Union, Regional Development Fund, the Mobility Plus programme of the Ministry of Science and Higher Education, the National Science Center (Poland), contracts Harmonia 2014/14/M/ST2/00428, Opus 2013/11/B/ST2/04202, 2014/13/B/ST2/02543 and 2014/15/B/ST2/03998, Sonata-bis 2012/07/E/ST2/01406; the Thalis and Aristeia programmes cofinanced by EU-ESF and the Greek NSRF; the National Priorities Research Program by Qatar National Research Fund; the Programa Clar\'in-COFUND del Principado de Asturias; the Rachadapisek Sompot Fund for Postdoctoral Fellowship, Chulalongkorn University and the Chulalongkorn Academic into Its 2nd Century Project Advancement Project (Thailand); and the Welch Foundation, contract C-1845.
\end{acknowledgments}

\bibliography{auto_generated}

\cleardoublepage \appendix\section{The CMS Collaboration \label{app:collab}}\begin{sloppypar}\hyphenpenalty=5000\widowpenalty=500\clubpenalty=5000\textbf{Yerevan Physics Institute,  Yerevan,  Armenia}\\*[0pt]
V.~Khachatryan, A.M.~Sirunyan, A.~Tumasyan
\vskip\cmsinstskip
\textbf{Institut f\"{u}r Hochenergiephysik,  Wien,  Austria}\\*[0pt]
W.~Adam, E.~Asilar, T.~Bergauer, J.~Brandstetter, E.~Brondolin, M.~Dragicevic, J.~Er\"{o}, M.~Flechl, M.~Friedl, R.~Fr\"{u}hwirth\cmsAuthorMark{1}, V.M.~Ghete, C.~Hartl, N.~H\"{o}rmann, J.~Hrubec, M.~Jeitler\cmsAuthorMark{1}, A.~K\"{o}nig, I.~Kr\"{a}tschmer, D.~Liko, T.~Matsushita, I.~Mikulec, D.~Rabady, N.~Rad, B.~Rahbaran, H.~Rohringer, J.~Schieck\cmsAuthorMark{1}, J.~Strauss, W.~Waltenberger, C.-E.~Wulz\cmsAuthorMark{1}
% ...
\vskip\cmsinstskip
% ...
1:~~Also at Vienna University of Technology, Vienna, Austria\\
% ...

\end{sloppypar}
\end{document}